\documentclass[aps,nofootinbib,preprintnumbers,showpacs,prd,twocolumn,superscriptaddress]{revtex4-1}
\usepackage{array}
\usepackage{booktabs}
\usepackage{subfigure}
\usepackage{capt-of}
\usepackage{graphicx}
\usepackage{lipsum}
\usepackage{epstopdf}
\usepackage{amsmath}
\usepackage{amssymb}
\usepackage{color,xcolor}
\usepackage[bookmarks=false]{hyperref}
\hypersetup{colorlinks=true,citecolor=green,linkcolor=blue,urlcolor=blue,pdfstartview=FitH,bookmarksopen=true}

\begin{document}
	
	\title{Generalized parton distributions of sea quark at zero skewness in the light-cone Model}
	\author{Xiaoyan Luan}
	\author{Zhun Lu}\email[]{zhulu@seu.edu.cn}
	
	\affiliation{Department of Physics, Southeast University, Nanjing 211189, China}

\begin{abstract}
We study the chiral-even generalized parton distributions (GPDs) of $\bar{u}$ and $\bar{d}$ quarks at zero skewness using the overlap representation within the light cone formalism.
The GPDs of $\bar{u}$ and $\bar{d}$ quarks can be expressed as the convolution of the light cone wave functions which are obtained from the baryon-meson fluctuation model in terms of the $|q\bar{q}B\rangle$ Fock states.
We present the numerical results for $H^{\bar{u}/P}(x,\xi,\Delta^2)$, $H^{\bar{d}/P}(x,\xi,\Delta^2)$, $E^{\bar{u}/P}(x,\xi,\Delta^2)$ and $E^{\bar{d}/P}(x,\xi,\Delta^2)$.
We apply the model resulting GPDs to calculate the orbital angular momentum of the $\bar{u}$ and $\bar{d}$ quarks, showing that $L^{\bar{u}/P}$, $L^{\bar{d}/P}$ are positive and $L^{\bar{u}/P}$ is smaller than $L^{\bar{d}/P}$.
The  sea quark OAM distributions in the impact parameter space $L_{\bar{q}}(x,\boldsymbol{b_T})$ are also calculated.
\end{abstract}
\maketitle
	
\section{Introduction}\label{Sec1}

The generalized parton distributions (GPDs)~\cite{Muller:1994ses,Ji:1996nm,Radyushkin:1997ki,Diehl:2015uka}, usually viewed as the extension of the standard parton distribution functions (PDFs), have been recognized as important quantities describing the three-dimensional structure of the nucleon in addition to the transverse momentum dependent parton distributions (TMDs).
As the Fourier transformation of the nonforward matrix elements of nonlocal operators, the GPDs appear in various exclusive processes, such as Deeply Virtual Compton Scattering (DVCS)~\cite{Ji:1996nm,Radyushkin:1996nd} and hard exclusive meson production (HEMP) \cite{Polyakov:1998ze,Collins:1996fb,Goloskokov:2008ib}.
At leading twist accuracy, there are eight GPDs: four of them are chiral-even: $H$, $E$, $\widetilde{H}$, $\widetilde{E}$; while the other four GPDs $H_T$, $E_T$, $\widetilde{H}_T$, $\widetilde{E}_T$ are chiral-odd.
The GPDs depend on three independent kinematic variables, the longitudinal momentum faction $x$ of the parton, the square of the total momentum transferred $t$ and the longitudinal momentum transferred skewness $\xi$.
In the forward limit, the GPDs reduce to the standard PDFs. On the other hand, the moments (integration over $x$) of the GPDs correspond to different form factors.
Thus they contain a wealth of information about the partonic structure of the hadron.
Particularly, the GPDs encode richer knowledge on the spin and orbital angular momentum (OAM) of the quarks and gluon inside the nucleon~\cite{Radyushkin:1997ki,Ji:1996nm,Sehgal:1974rz,Kroll:2020jat} than the standard PDFs do.
Through a Fourier transformation with respect to the transverse momentum transfer $\Delta_T$, one can obtain the distributions in the impact parameter space which provides tomographic description of the nucleon structure.

In recent years, a lot of experimental and theoretical studies related to GPDs have been carried out. The experimental data on DVCS and HEMP were collected by the H1 collaboration~\cite{H1:1999pji,H1:2001nez,H1:2005gdw}, the ZEUS collaboration \cite{ZEUS:1998xpo,ZEUS:2003pwh}, well as the fixed target experiments at HERMES \cite{HERMES:2001bob,HERMES:2011bou,HERMES:2012gbh}, COMPASS~\cite{dHose:2004usi}.
and JLab~\cite{CLAS:2001wjj}.
In addition to experimental measurements, considerable progress has been made in theoretical studies.
Due to their nonperturbative nature, the GPDs cannot be calculated directly from the first principle of QCD.
Although some breakthroughs have been made in the simulation of GPDs on the lattice, it is still in the early stage of development~\cite{Ji:2013dva,Orginos:2017kos,Ma:2014jla,Ma:2017pxb}.
Therefore, we still rely on models in order to obtain useful information on GPDs.
Various model calculations have been applied to calculate the GPDs, such as the MIT bag model~\cite{Ji:1997gm}, the (light-front) constituent quark model ~\cite{boffi2003linking,Scopetta:2003et,Choi:2001fc,Choi:2002ic}, the NJL model~\cite{Mineo:2005vs}, the color glass condensate model~\cite{Goeke:2008jz}, the chiral quark-soliton model~\cite{goeke2001hard,Ossmann:2004bp}, the Bethe-Salpeter approach~\cite{Tiburzi:2001je,Theussl:2002xp}, and the meson cloud model~\cite{Pasquini:2006dv,pasquini2007generalized}.
Among them, a complete overlap representation of GPDs has also been worked out within the light cone formalism~\cite{Diehl:2000xz,Brodsky:2000xy}.
In this approach, the Fock-state expansion of hadron is performed in terms of $N$-parton Fock states with coefficients representing the light-cone wavefunction (LCWF) of the $N$ partons~\cite{Brodsky:2000xy,Muller:2014tqa,Pasquini:2006dv}.
Within this approach, the GPDs have a simple interpretation in terms of LCWFs by the overlap representation~\cite{Brodsky:2000xy,Muller:2014tqa}.
	
In this paper, we apply the light-cone quark model to calculate the chiral-even GPDs of the $\bar{u}$ and $\bar{d}$ quarks at zero skewness using the overlap representation.
As proposed in Refs.~\cite{Brodsky:1996hc,Luan:2022fjc}, the sea quark degree freedom is generated by the assumption that the proton can fluctuate to a composite state containing a meson $M$ and a baryon $B$, which is similar to the meson cloud effect proposed by Sullivan~\cite{Sullivan:1971kd}.
The LCWFs of the proton can be obtained in terms of the $|q\bar{q}B\rangle$ Fock states, where $q\bar{q}$ are components of pion meson.
In this framework, the GPDs of $\bar{q}$ (for examplem, $H^{\bar{q}/P}$) can be expressed as the convolution of the GPDs of the pion inside the proton ($H^{\pi/P}$) and the GPDs of $\bar{q}$ inside the pion ($H^{\bar{q}/\pi}$).
This form of convolution consists with those in Refs.~\cite{Scopetta:2006wt,Pasquini:2006dv,He:2022leb}.
As a check, we compare our numerical result with previous model calculation~\cite{He:2022leb}, which adopts the nonlocal chiral effective theory on the GPDs of sea quark in the proton.
An important implication of GPDs is that they are related to the angular momentum of the parton.
Thereby, we apply the model resulting GPDs to estimate the sea quark OAM as well as study the impact parameter dependence of the sea quark OAM distribution $L^{\bar{q}/P}(x, b_T )$.
	
The rest part of the paper is organized as follows.
In Sec. II, we apply the LCWFs motivated by the baryon-meson fluctuation model to obtain the analytical expressions of the three GPDs of the sea quarks.
In Sec. III, we present the numerical results of these GPDs and the OAM contributed by the sea quarks.  The model results of the sea quark OAM distributions and their impact-parameter dependence are also provided. We summarize the paper in Sec. VI.

\section{the chiral-even GPDs of the sea quark}\label{Sec2}

The light-cone formalism provides a convenient way to calculate the GPDs~\cite{Brodsky:2000xy}.
In this approach, the LCWFs of the proton are obtained in terms of a hadronic composite state in Fock-state basis.
Recently, the overlap representation has been applied to calculate the GPDs using the LCWFs ~\cite{Brodsky:2000xy,Muller:2014tqa,Burkardt:2003je}.
In this section, we will calculate the chiral-even of sea quark at zero skewness via the overlap representation within the light cone formalism.
	
For the generation of the sea quark degree of freedom, we apply the baryon-meson fluctuation model~\cite{Brodsky:1996hc}, in which the proton can fluctuate to a composite system formed by a meson $M$ and a baryon $B$, where the meson is composed in terms of $q\bar{q}$.
\begin{align}\label{fock state}
|p\rangle\to| M B\rangle\to|q\bar{q}B\rangle.
\end{align}
For our purpose we consider the fluctuation $|p\rangle \to |\pi^+ n\rangle $ and $|p\rangle \to |\pi^- \Delta^{++}\rangle $.
The details of the model can be found in Ref.~\cite{Brodsky:1996hc,Luan:2022fjc}.
For the above proton composite state, the LCWFs have also been derived in Ref.~\cite{Luan:2022fjc}.
\begin{align}\label{LCWFs}
\psi^{\lambda_N}_{{\lambda_B}{\lambda_q}{\lambda_{\bar{q}}}}(x,y,\boldsymbol{k}_T,\boldsymbol{r}_T)
=&\psi^{\lambda_N}_{\lambda_B}(y,\boldsymbol{r}_T)\psi_{{\lambda_q}{\lambda_{\bar{q}}}}
(x,y,\boldsymbol{k}_T,\boldsymbol{r}_T),
\end{align}
where $\psi^{\lambda_N}_{\lambda_B}(y,r_T)$ can be viewed as the wave function of the nucleon in terms $\pi B$ components, and $\psi_{{\lambda_q}{\lambda_{\bar{q}}}}(x,y,\boldsymbol{k}_T,\boldsymbol{r}_T)$ is the pion LCWFs in terms of $q \bar{q}$ components.
The indices $\lambda_N$, $\lambda_B$, $\lambda_q$, $\lambda_{\bar{q}}$ denote the helicity of the proton, the baryon, the quark and sea quark, respectively.
$x$ and $y$ represent their light-cone momentum fractions, $\boldsymbol{k}_T$ and $\boldsymbol{r}_T$ denote the transverse momenta of the antiquark and the meson.
For the former one of Eq.~(\ref{LCWFs}), they have the expression:
\begin{align}\label{former}
    	\notag\psi^+_+(y,\boldsymbol{r}_T)&=\frac{M_B-(1-y)M}{\sqrt{1-y}}\phi_1, \\
    	\notag\psi^+_-(y,\boldsymbol{r}_T)&=\frac{r_1+ir_2}{\sqrt{1-y}}\phi_1, \\
    	\notag\psi^-_+(y,\boldsymbol{r}_T)&=\frac{r_1-ir_2}{\sqrt{1-y}}\phi_1 , \\
    	\psi^-_-(y,\boldsymbol{r}_T)&=\frac{(1-y)M-M_B}{\sqrt{1-y}}\phi_1.
\end{align}	
Here, $M$ and $M_B$ are the masses of proton and baryon, respectively.
$\phi_1$ is the wave function of the baryon-meson system in the momentum space with the form
\begin{align}
\phi_1(y,\boldsymbol{r}_T)=-\frac{g(r^2)\sqrt{y(1-y)}}{\boldsymbol{r}_T^2+L_1^2(m_\pi^2)},
\end{align}
where $m_\pi$ is the mass of $\pi$ meson, $g(r^2)$ is the form factor for the coupling of the nucleon-pion-baryon vertex, and
\begin{align}
    	L_1^2({m_\pi^2})=yM_B^2+(1-y){m_\pi^2}-y(1-y)M^2.
\end{align}
Finally, the wave functions of the pion meson in terms of the $q \bar{q}$ pair in Eq.~(\ref{LCWFs}) have the following expressions:
\begin{align}\label{later}
\notag\psi{_+}{_+}(x,y,\boldsymbol{k}_T,\boldsymbol{r}_T)&=\frac{my}{\sqrt{x(y-x)}}\phi_2,\\	\notag\psi{_+}{_-}(x,y,\boldsymbol{k}_T,\boldsymbol{r}_T)
&=\frac{y(k_1-ik_2)-x(r_1-ir_2)}{\sqrt{x(y-x)}}\phi_2, \\ \notag\psi{_-}{_+}(x,y,\boldsymbol{k}_T,\boldsymbol{r}_T)
&=\frac{y(k_1+ik_2)-x(r_1+ir_2)}{\sqrt{x(y-x)}}\phi_2, \\
\psi{_-}{_-}(x,y,\boldsymbol{k}_T,\boldsymbol{r}_T)&=\frac{-my}{\sqrt{x(y-x)}}\phi_2,
\end{align}
with $m$ the mass of quark and the sea quark.
Again, $\phi_2$ is the wave function of the pion meson in momentum space
\begin{align}	     	
\phi_2(x,y,\boldsymbol{k}_T,\boldsymbol{r}_T)
=-\frac{g(k^2)\sqrt{\frac{x}{y}(1-\frac{x}{y})}}
{(\boldsymbol{k}_T-\frac{x}{y}\boldsymbol{r}_T)^2+L_2^2(m^2)},
\end{align}
$g(k^2)$ is the form factor for the coupling of the pion meson-quark-sea quark vertex, and
\begin{align}
L_2^2(m^2)=\frac{x}{y}m^2+\left(1-\frac{x}{y}\right)m^2-\frac{x}{y}\left(1-\frac{x}{y}\right){m_\pi}^2.
\end{align}
In our calculation we adopt the dipolar form for $g(r^2)$ and $g(k^2)$
\begin{align}
g(r^2)&=-g_1(1-y)\frac{\boldsymbol{r}_T^2+L_1^2(m_\pi^2)}
{[\boldsymbol{r}_T^2+L_1^2(\Lambda^2_\pi)]^2},\label{eq15}	\\
g(k^2)&=-g_2(1-\frac{x}{y})\frac{(\boldsymbol{k}_T-\frac{x}{y}\boldsymbol{r}_T)^2+L_2^2(m^2)}
{[(\boldsymbol{k}_T-\frac{x}{y}\boldsymbol{r}_T)^2+L_2^2(\Lambda^2_{\bar{q}})]^2}.\label{eq16}	\end{align}
where $g_1$ and $g_2$ are the free parameters representing the couplings of the vertices.

With those LCWFs, we can calculate the chiral-even GPDs at zero skewness.
In the overlap representation~\cite{Brodsky:2000xy,Muller:2014tqa}, the GPD $H$ at $\xi=0$ can be expressed as
\begin{align}\label{GPDH}
\notag 	&H(x,y,0,-\boldsymbol{\Delta}_T^2)=
\nonumber\\&\sum_{\{\lambda\}}\int\frac{d^2\boldsymbol{k}_T}{16\pi^3}
\int\frac{d^2\boldsymbol{r}_T}{16\pi^3}\psi^{\uparrow*}_{\{\lambda\}}
(x,y,\boldsymbol{k}^{{\prime}{\prime}}_T,\boldsymbol{r}^{{\prime}{\prime}}_T)
\psi^{\uparrow}_{\{\lambda\}}(x,y,\boldsymbol{k}^{\prime}_T,\boldsymbol{r}^{\prime}_T)
\nonumber \\ \notag&=\int\frac{d^2\boldsymbol{r}_T}{16\pi^3}\sum_{\lambda_B}
\psi^{\uparrow*}_{\lambda_B}(y,\boldsymbol{r}^{{\prime}{\prime}}_T)
\psi^{\uparrow}_{\lambda_B}(y,\boldsymbol{r}^{\prime}_T)  \nonumber\\ &\times\int\frac{d^2\boldsymbol{k}_T}{16\pi^3}\sum_{{\lambda_q}{\lambda_{\bar{q}}}}
\psi^*_{{\lambda_q}{\lambda_{\bar{q}}}}(x,y,\boldsymbol{k}^{{\prime}{\prime}}_T,
\boldsymbol{r}^{{\prime}{\prime}}_T)\psi_{{\lambda_q}{\lambda_{\bar{q}}}}
(x,y,\boldsymbol{k}^{\prime}_T,\boldsymbol{r}^{\prime}_T)\nonumber\\
&=H^{\pi/P}(y,0,-\boldsymbol{\Delta}_T^2)H^{\bar{q}/\pi}(\frac{x}{y},0,-\boldsymbol{\Delta}_T^2),
\end{align}
where $\{\lambda\}={\lambda_N,\lambda_B, \lambda_q, \lambda_{\bar{q}}}$, with
\begin{align}
\nonumber\boldsymbol{k}^{{\prime}{\prime}}_T&=\boldsymbol{k}_T-\frac{1}{2}(1-x)\boldsymbol{\Delta}_T  \\ \boldsymbol{k}^{\prime}_T&=\boldsymbol{k}_T+\frac{1}{2}(1-x)\boldsymbol{\Delta}_T
\end{align}	
for the final and initial struck quark $\bar{q}$, and
\begin{align}
\nonumber-\boldsymbol{r}^{{\prime}{\prime}}_T&=-\boldsymbol{r}_T+\frac{1}{2}(1-y)\boldsymbol{\Delta}_T   \\ \nonumber-\boldsymbol{r}^{\prime}_T&=-\boldsymbol{r}_T-\frac{1}{2}(1-y)\boldsymbol{\Delta}_T\\ \nonumber(\boldsymbol{r}_T-\boldsymbol{k}_T)^{{\prime}{\prime}}&=	(\boldsymbol{r}_T-\boldsymbol{k}_T)+\frac{1}{2}(y-x)\boldsymbol{\Delta}_T\\
(\boldsymbol{r}_T-\boldsymbol{k}_T)^{\prime}&=	(\boldsymbol{r}_T-\boldsymbol{k}_T)-\frac{1}{2}(y-x)\boldsymbol{\Delta}_T
\end{align}	
for the final and initial spectators $\pi$ and $q$.

In the above Eq.~(\ref{GPDH}), $H^{\pi/P}(y,0,-\boldsymbol{\Delta}_T^2)$ is the electric GPD of the pion inside the proton:
\begin{align}\label{piGPDH}
\notag &H^{\pi/P}(y,0,-\boldsymbol{\Delta}_T^2)=\frac{g^2_1y(1-y)^2}{16\pi^3}
\\&\times\int d^2\boldsymbol{r}_T\frac{\boldsymbol{r}_T^2-\frac{1}{4}(1-y)^2\boldsymbol{\Delta}_T^2
+[M_B-(1-y)M]^2}{D_1(y,\boldsymbol{r}_T,\boldsymbol{\Delta}_T)},
\end{align}
where
\begin{align} 	
&D_1(y,\boldsymbol{r}_T,\boldsymbol{\Delta}_T)	=[(\boldsymbol{r}_T-\frac{1}{2}(1-y)\boldsymbol{\Delta}_T)^2+L_1^2]^2\nonumber\\
&\times[(\boldsymbol{r}_T+\frac{1}{2}(1-y)\boldsymbol{\Delta}_T)^2+L_1^2]^2,
\end{align}
and $H^{\bar{q}/\pi}(\frac{x}{y},0,-\boldsymbol{\Delta}_T^2)$ denotes the GPD for the anti-quark inside the pion with the momentum fraction $x/y$:
\begin{align}\label{qbarGPDH}
\nonumber &H^{\bar{q}/\pi}(\frac{x}{y},0,-\boldsymbol{\Delta}_T^2)
\\&=\frac{g_2^2(1-\frac{x}{y})^2}{8\pi^3}\int d^2\boldsymbol{k}_T\frac{(\boldsymbol{k}_T-\frac{x}{y}\boldsymbol{r}_T)^2
-\frac{1}{4}(1-\frac{x}{y})^2\boldsymbol{\Delta}_T^2+m^2}{  D_2(\frac{x}{y},\boldsymbol{k}_T-\frac{x}{y}\boldsymbol{r}_T,\boldsymbol{\Delta}_T)},
\end{align}
where
\begin{align}
\nonumber &D_2(\frac{x}{y},\boldsymbol{k}_T-\frac{x}{y}\boldsymbol{r}_T,\boldsymbol{\Delta}_T)\\
\nonumber&=\left[[(\boldsymbol{k}_T-\frac{x}{y}\boldsymbol{r}_T)-\frac{1}{2}(1-\frac{x}{y})
\boldsymbol{\Delta}_T]^2+L_2^2\right]^2
\\&\times\left[[(\boldsymbol{k}_T-\frac{x}{y}\boldsymbol{r}_T)
+\frac{1}{2}(1-\frac{x}{y})\boldsymbol{\Delta}_T]^2+L_2^2\right]^2.
\end{align}
After integrating out the light-cone momentum fraction $y$, we can obtain the GPD of the sea quark inside the proton
\begin{align}	
H^{\bar{q}/P}(x,0,-\boldsymbol{\Delta}_T^2)
=&\int_{x}^{1}\frac{dy}{y}H^{\pi/P}(y,0,-\boldsymbol{\Delta}_T^2)\nonumber\\
\times&H^{\bar{q}/\pi}(\frac{x}{y},0,-\boldsymbol{\Delta}_T^2).
\end{align}
Its final expression can be written as
\begin{align}\nonumber &H^{\bar{q}/P}(x,0,-\boldsymbol{\Delta}_T^2)=\frac{g^2_1g_2^2}{2(2\pi)^6}\int_{x}^{1}\frac{dy}{y}\int d^2\boldsymbol{k}_T\int d^2\boldsymbol{r}_T\\
\nonumber&\times\frac{y(1-y)^2\left[\boldsymbol{r}_T^2
-\frac{1}{4}(1-y)^2\boldsymbol{\Delta}_T^2+[M_B-(1-y)M]^2\right]}
{D_1(y,\boldsymbol{r}_T,\boldsymbol{\Delta}_T)}\\
&\times\frac{(1-\frac{x}{y})^2\left[(\boldsymbol{k}_T-\frac{x}{y}\boldsymbol{r}_T)^2
-\frac{1}{4}(1-\frac{x}{y})^2\boldsymbol{\Delta}_T^2+m^2\right]}{  D_2(\frac{x}{y},\boldsymbol{k}_T-\frac{x}{y}\boldsymbol{r}_T,\boldsymbol{\Delta}_T)}.
\end{align}

\begin{figure*}[htbp]
    	\centering
    	\subfigure{\begin{minipage}[b]{0.45\linewidth}
    			\centering
    			\includegraphics[width=\linewidth]{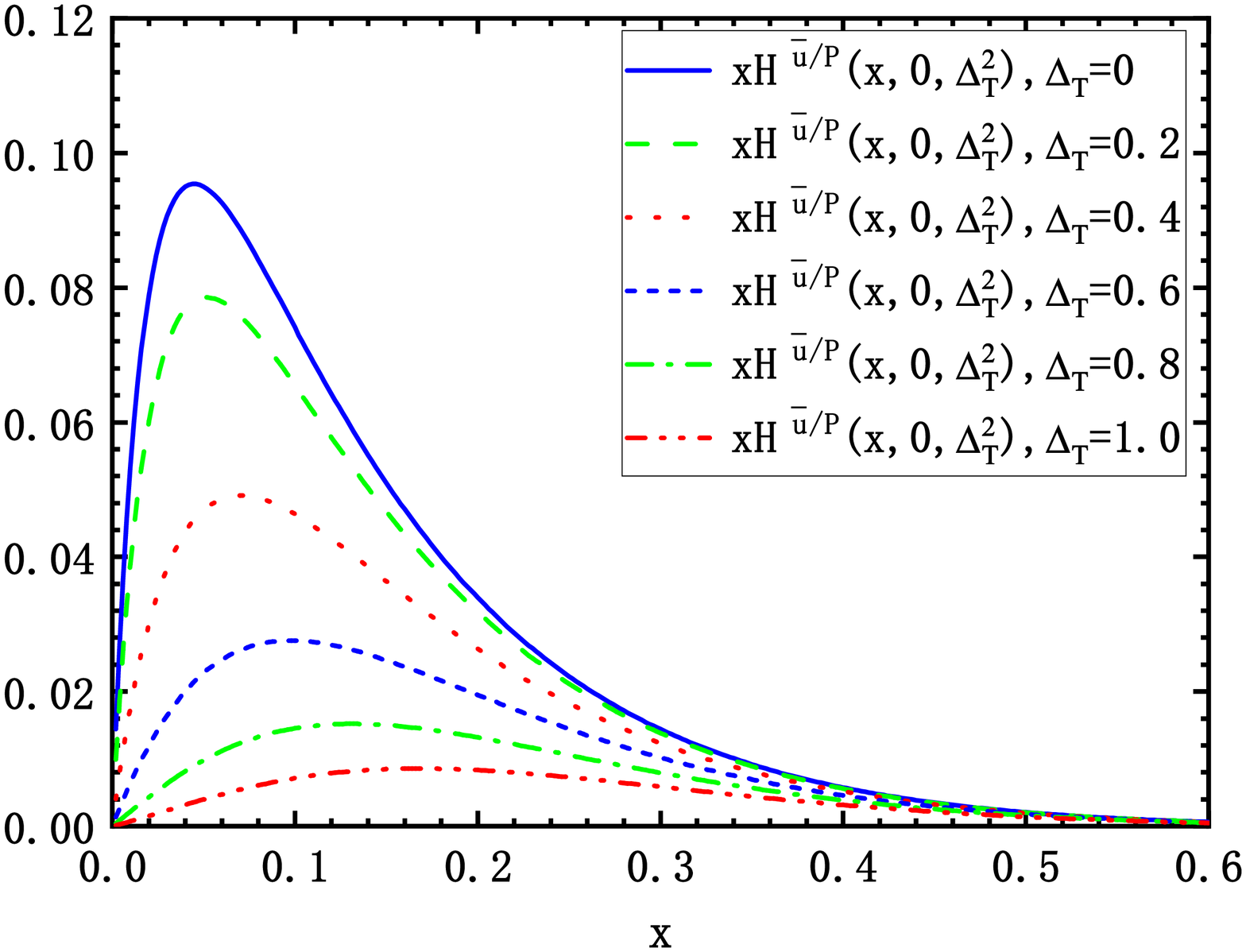}
    	\end{minipage}}
    	\subfigure{\begin{minipage}[b]{0.45\linewidth}
    			\centering
    			\includegraphics[width=\linewidth]{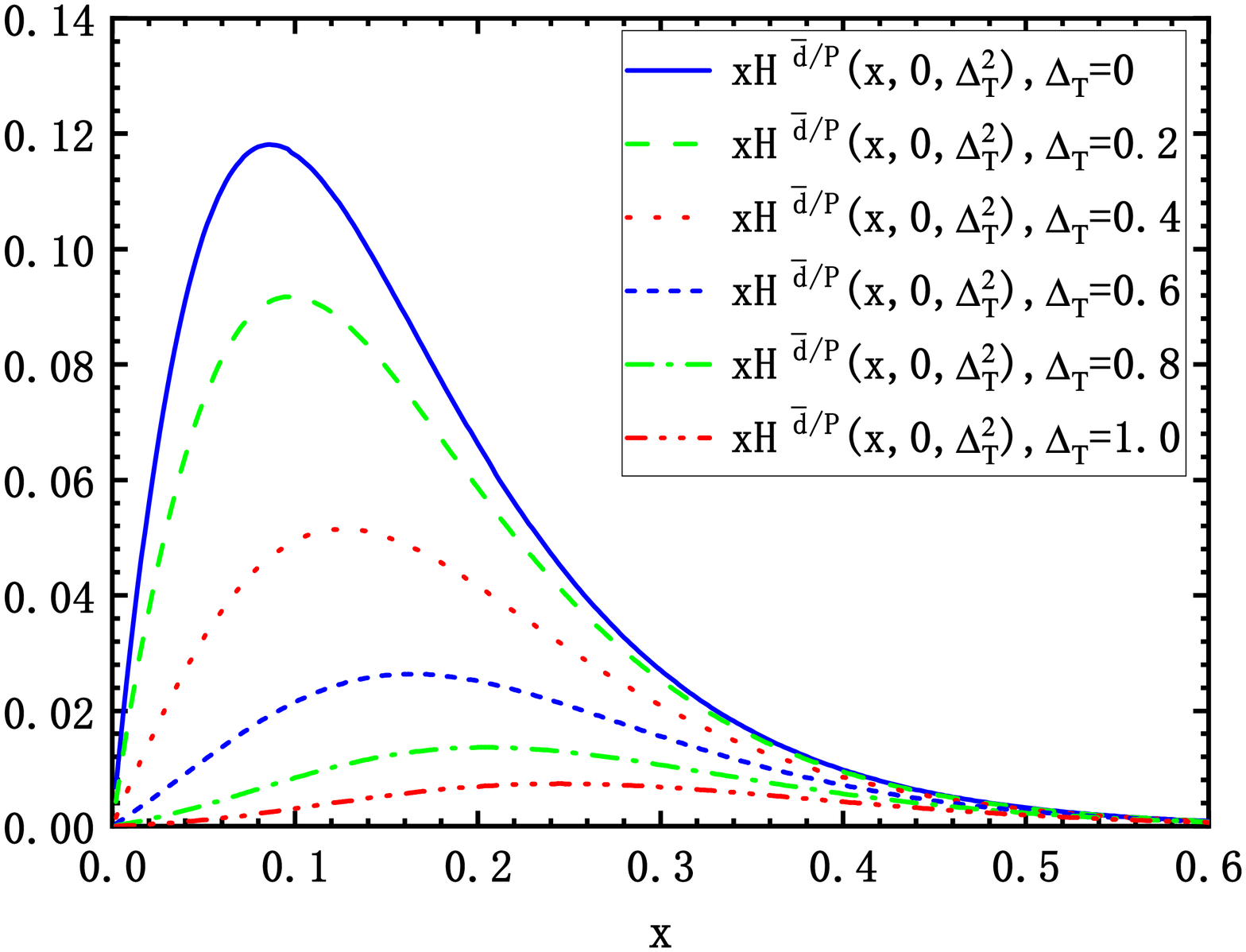}    	
      \end{minipage}}
    \caption{The GPDs $H^{\bar{u}/P}(x,0,-\boldsymbol{\Delta}_T^2)$ and $H^{\bar{d}/P}(x,0,-\boldsymbol{\Delta}_T^2)$ for the proton in the light-cone quark model as functions of $x$ at different values of $\Delta_T$ .} \label{H}
\end{figure*}

Similarly, the magnetic GPD of the sea quark $E^{\bar{q}/P}$ can be calculated from the overlap representation of the LCWFs:
\begin{align}\label{GPDE}\notag
&\frac{\Delta_1-i\Delta_2}{2M}E^{\bar{q}/P}(x,y,0,-\boldsymbol{\Delta}_T^2)=\\
\nonumber&\sum_{\{\lambda\}}\int\frac{d^2\boldsymbol{k}_T}{16\pi^3}
\int\frac{d^2\boldsymbol{r}_T}{16\pi^3}\psi^{\uparrow*}_{\{\lambda\}}
(x,y,\boldsymbol{k}^{{\prime}{\prime}}_T,\boldsymbol{r}^{{\prime}{\prime}}_T)
\psi^{\downarrow}_{\{\lambda\}}(x,y,\boldsymbol{k}^{\prime}_T,\boldsymbol{r}^{\prime}_T)\\
\notag&=\int\frac{d^2\boldsymbol{r}_T}{16\pi^3}\sum_{\lambda_B}
\psi^{\uparrow*}_{\lambda_B}(y,\boldsymbol{r}^{{\prime}{\prime}}_T)
\psi^{\downarrow}_{\lambda_B}(y,\boldsymbol{r}^{\prime}_T)\\
\nonumber&\times\int\frac{d^2\boldsymbol{k}_T}{16\pi^3}\sum_{{\lambda_q}
{\lambda_{\bar{q}}}}\psi^*_{{\lambda_q}{\lambda_{\bar{q}}}}(x,y,\boldsymbol{k}^{{\prime}{\prime}}_T,
\boldsymbol{r}^{{\prime}{\prime}}_T)\psi_{{\lambda_q}{\lambda_{\bar{q}}}}
(x,y,\boldsymbol{k}^{\prime}_T,\boldsymbol{r}^{\prime}_T)	\\&=\frac{\Delta_1-i\Delta_2}{2M}E^{\pi/P}(y,0,-\boldsymbol{\Delta}_T^2)
H^{\bar{q}/\pi}(\frac{x}{y},0,-\boldsymbol{\Delta}_T^2).
\end{align}
In the above equation, we use $E^{\pi/P}(y,0,-\boldsymbol{\Delta}_T^2)$ to denote the magnetic GPD of the pion inside the proton, and in our model it may be easily calculated from the LCWFs of the proton in terms of $\pi B$ component:
\begin{align}\label{piGPDE}
&E^{\pi/P}(y,0,-\boldsymbol{\Delta}_T^2)\nonumber\\
=&\frac{g_1^2y(1-y)^3}{8\pi^3}\int d^2\boldsymbol{r}_T\frac{M[M_B-(1-y)M]}{D_1(y,\boldsymbol{r}_T,\boldsymbol{\Delta}_T)},
\end{align}
Again, after integrating out the light-cone momentum fraction $y$, we can obtain $E^{\bar{q}/P}(x,0,-\boldsymbol{\Delta}_T^2)$
\begin{align}	
E^{\bar{q}/P}(x,0,-\boldsymbol{\Delta}_T^2)
=&\int_{x}^{1}\frac{dy}{y}E^{\pi/P}(y,0,-\boldsymbol{\Delta}_T^2)\nonumber\\
\times&H^{\bar{q}/\pi}(\frac{x}{y},0,-\boldsymbol{\Delta}_T^2),
\end{align}
and its full expression has the form
\begin{align}\label{qbarGPDE}\notag &E^{\overline{q}/P}(x,0,-\boldsymbol{\Delta}_T^2)=\frac{g^2_1g_2^2}{(2\pi)^6}\int_{x}^{1}\frac{dy}{y}\int d^2\boldsymbol{k}_T\int d^2\boldsymbol{r}_T\\
&\times\frac{y(1-y)^3M[M_B-(1-y)M]}{D_1(y,\boldsymbol{r}_T,\boldsymbol{\Delta}_T)}\nonumber\\
&\times\frac{(1-\frac{x}{y})^2\left[(\boldsymbol{k}_T-\frac{x}{y}\boldsymbol{r}_T)^2
-\frac{1}{4}(1-\frac{x}{y})^2\boldsymbol{\Delta}_T^2+m^2\right]}{  D_2(\frac{x}{y},\boldsymbol{k}_T-\frac{x}{y}\boldsymbol{r}_T,\boldsymbol{\Delta}_T)}.
\end{align}

Similar to the case of the electric GPD and the magnetic GPD, $\widetilde{H}$ can be also calculated from the overlap representation of the LCWFs
\begin{align}
\notag 	&\widetilde{H}(x,y,0,-\boldsymbol{\Delta}_T^2)=\sum_{\{\lambda\}}sign(\lambda_{\bar{q}})\\
\nonumber&\times\int\frac{d^2\boldsymbol{k}_T}{16\pi^3}\int\frac{d^2\boldsymbol{r}_T}{16\pi^3}
\psi^{\uparrow*}_{\{\lambda\}}(x,y,\boldsymbol{k}^{{\prime}{\prime}}_T,\boldsymbol{r}^{{\prime}{\prime}}_T)
\psi^{\uparrow}_{\{\lambda\}}(x,y,\boldsymbol{k}^{\prime}_T,\boldsymbol{r}^{\prime}_T)	\\
\nonumber&=\int\frac{d^2\boldsymbol{r}_T}{16\pi^3}\sum_{\lambda_B}
\psi^{\uparrow*}_{\lambda_B}(y,\boldsymbol{r}^{{\prime}{\prime}}_T)
\psi^{\uparrow}_{\lambda_B}(y,\boldsymbol{r}^{\prime}_T)  \\
\nonumber&\times\int\frac{d^2\boldsymbol{k}_T}{16\pi^3}\sum_{{\lambda_q}{\lambda_{\bar{q}}}}
sign(\lambda_{\bar{q}})\psi^*_{{\lambda_q}{\lambda_{\bar{q}}}}(x,y,\boldsymbol{k}^{{\prime}{\prime}}_T,
\boldsymbol{r}^{{\prime}{\prime}}_T)\nonumber\\
&\times\psi_{{\lambda_q}{\lambda_{\bar{q}}}}(x,y,\boldsymbol{k}^{\prime}_T,
\boldsymbol{r}^{\prime}_T)	\nonumber\\
&=H^{\pi/P}(y,0,-\boldsymbol{\Delta}_T^2)\widetilde{H}^{\bar{q}/\pi}(\frac{x}{y},0,-\boldsymbol{\Delta}_T^2),
\end{align}
where $\widetilde{H}^{\bar{q}/\pi}(\frac{x}{y},0,-\boldsymbol{\Delta}_T^2)$ is the GPD for the anti-quark in the pion. Since
\begin{align}
\notag &\widetilde{H}^{\bar{q}/\pi}(\frac{x}{y},0,-\boldsymbol{\Delta}_T^2)	=\int\frac{d^2\boldsymbol{k}_T}{16\pi^3}\sum_{{\lambda_q}{\lambda_{\bar{q}}}}
sign(\lambda_{\bar{q}})
\\
\nonumber&\times\psi^*_{{\lambda_q}{\lambda_{\bar{q}}}}(x,y,\boldsymbol{k}^{{\prime}{\prime}}_T,
\boldsymbol{r}^{{\prime}{\prime}}_T)\psi_{{\lambda_q}{\lambda_{\bar{q}}}}
(x,y,\boldsymbol{k}^{\prime}_T,\boldsymbol{r}^{\prime}_T)\\&=0,
\end{align}
we find in our model
\begin{align}
\notag 	\widetilde{H}^{\bar{q}/P}(x,0,-\boldsymbol{\Delta}_T^2)&
=\int_{x}^{1}\frac{dy}{y}H^{\pi/P}(y,0,-\boldsymbol{\Delta}_T^2)\nonumber\\
&\times\widetilde{H}^{\bar{q}/\pi}(\frac{x}{y},0,-\boldsymbol{\Delta}_T^2)=0.
\end{align}

\section{Numerical results for the sea quark GPDs and OAM}\label{Sec4}

\begin{center}
    	\setlength{\tabcolsep}{5mm}
    	\renewcommand\arraystretch{1.5}
    	\begin{tabular}{ c | c | c }
    		\hline
    		Parameters & $\bar{u}$ & $\bar{d}$ \\
    		\hline
    		\hline
    		$g_1$ & 9.33 & 5.79 \\
    		\hline
    		$g_2$ & 4.46 & 4.46 \\
    		\hline
    		$\Lambda_\pi(GeV)$ &  0.223 & 0.223 \\
    		\hline
    		$\Lambda_{\bar{q}}(GeV)$ &  0.510 &  0.510 \\
    		\hline
    	\end{tabular}
    	\captionof{table}{Values of the parameters obtained from Ref.~\cite{Luan:2022fjc}.} \label{tab1}
      \end{center}

\begin{figure*}[htbp]
    	\centering
    	\subfigure{\begin{minipage}[b]{0.45\linewidth}
    			\centering
    			\includegraphics[width=\linewidth]{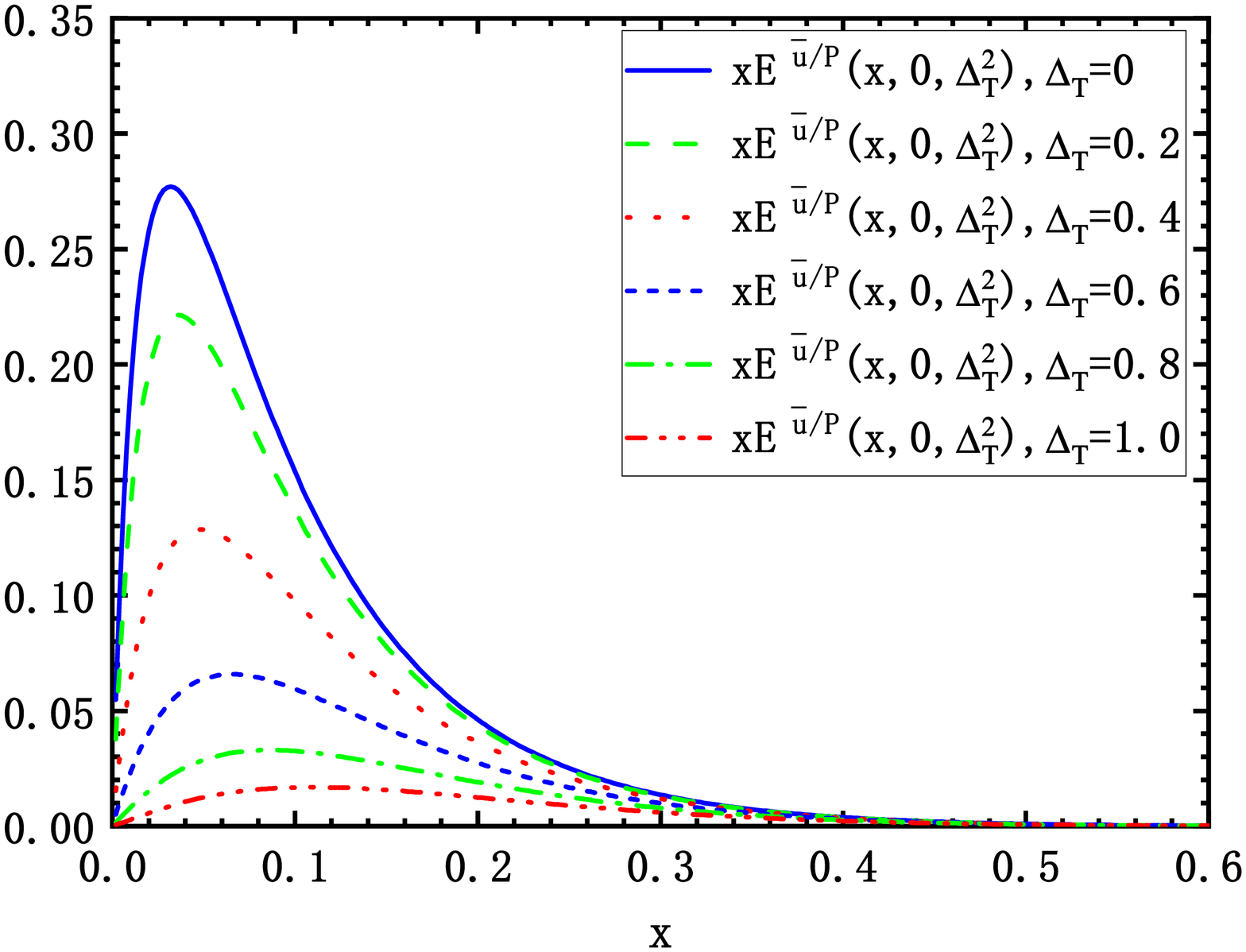}
    	\end{minipage}}
    	\subfigure{\begin{minipage}[b]{0.45\linewidth}
    			\centering
    			\includegraphics[width=\linewidth]{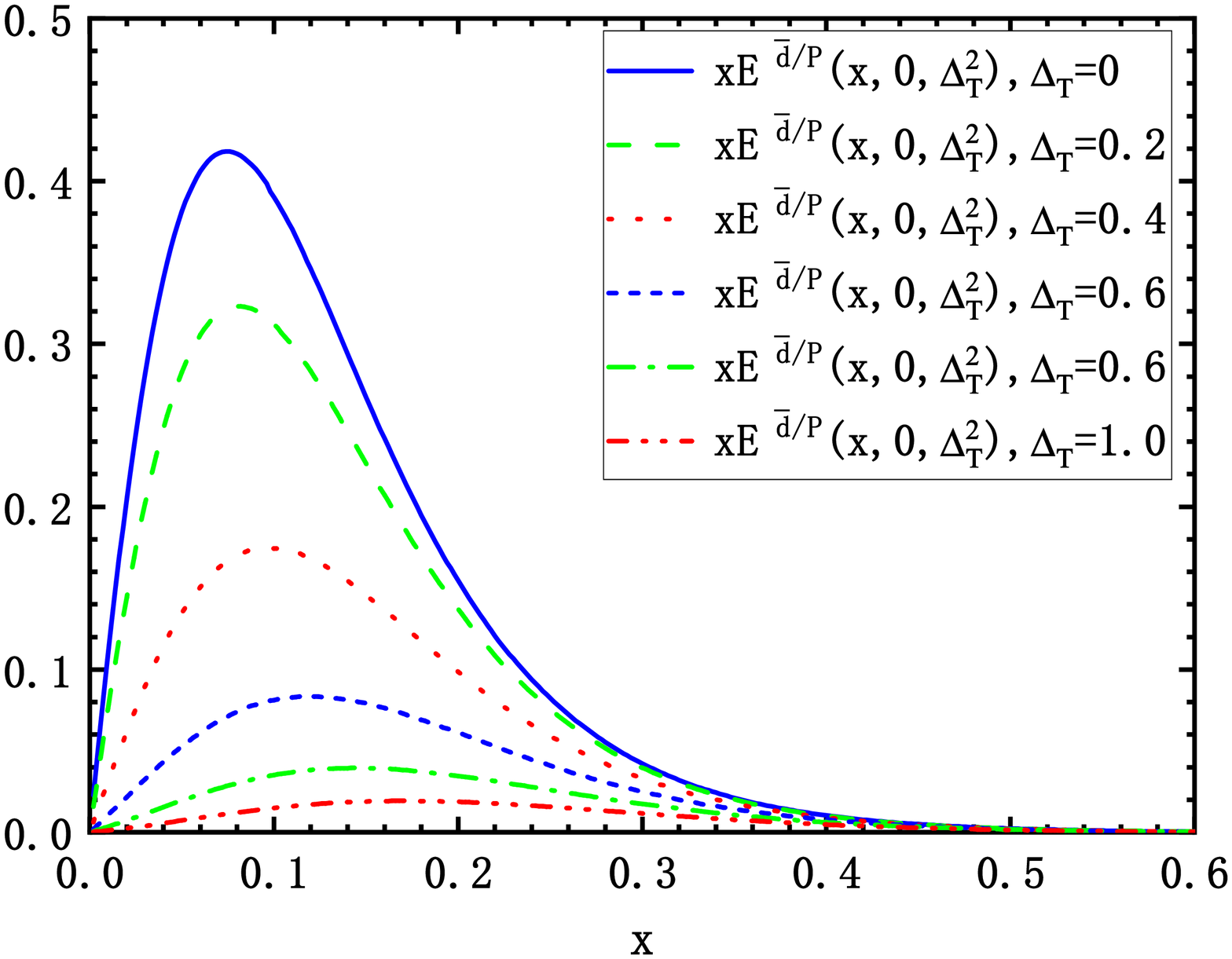}    	
    \end{minipage}}
    \caption{The GPDs $E^{\bar{u}/P}(x,0,-\boldsymbol{\Delta}_T^2)$ and $E^{\bar{d}/P}(x,0,-\boldsymbol{\Delta}_T^2)$ for the proton in the light-cone quark model as functions of $x$ at different values of $\Delta_T$.} \label{E}
\end{figure*}

\begin{figure*}[htbp]
    	\centering
    	\subfigure{\begin{minipage}[b]{0.45\linewidth}
    			\centering
    			\includegraphics[width=\linewidth]{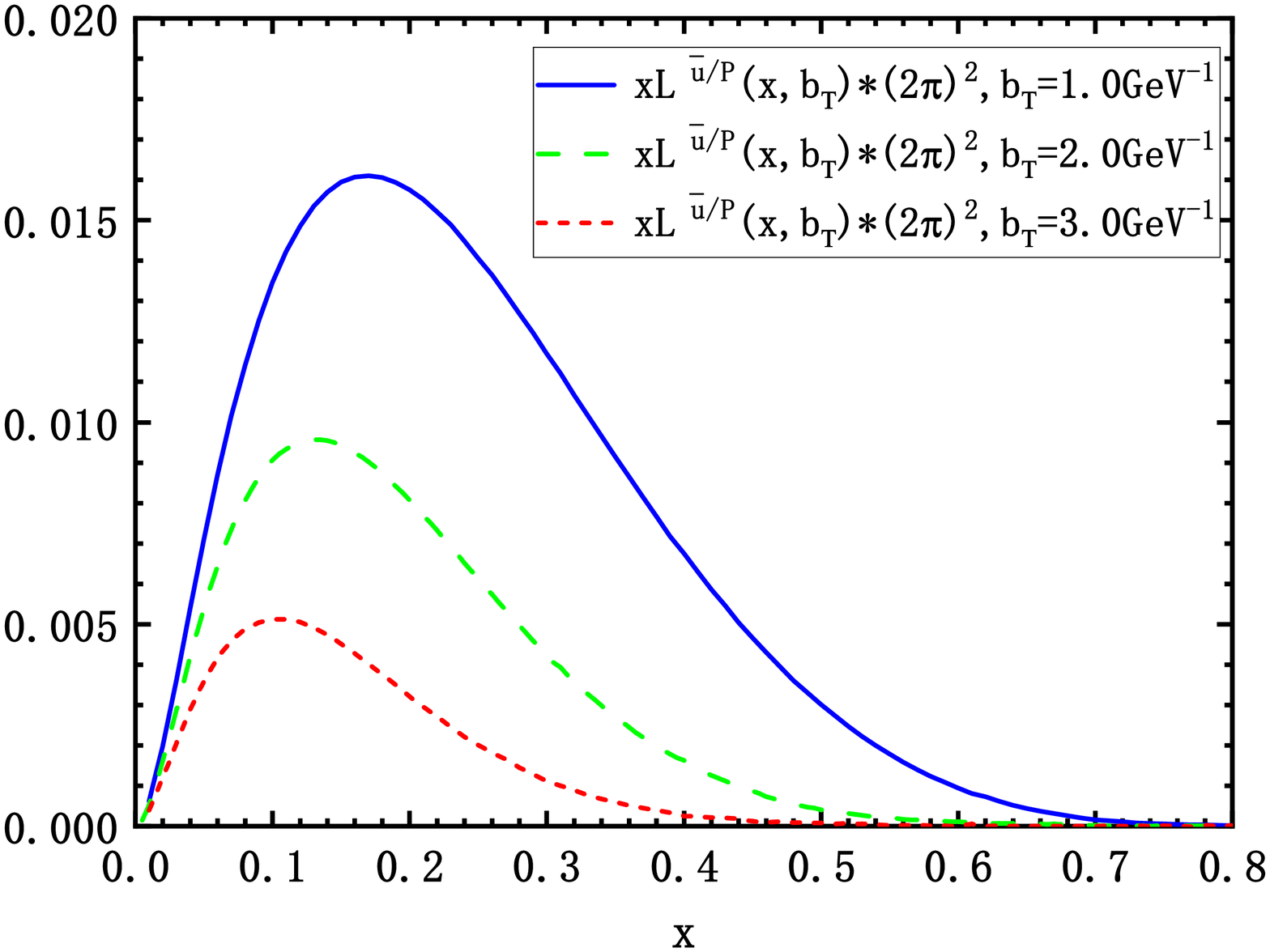}
    	\end{minipage}}
    	\subfigure{\begin{minipage}[b]{0.45\linewidth}
    			\centering
    			\includegraphics[width=\linewidth]{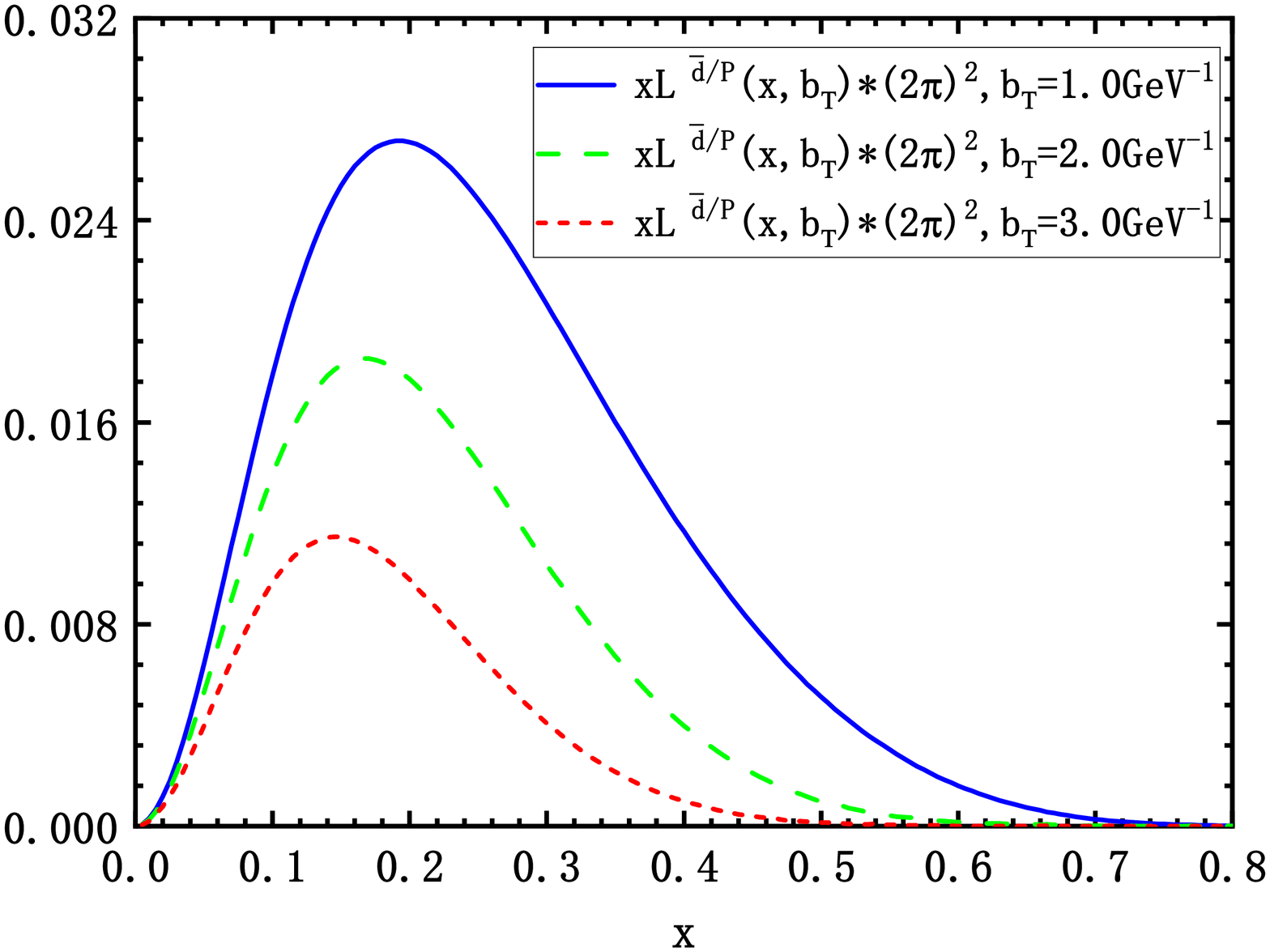}    	
       \end{minipage}}
       \caption{The impact parameter distributions (scaled with a factor of $(2\pi)^2$) $L^{\bar{u}/P} (x,\boldsymbol{b}_T^2)$ (left) and $L^{\bar{d}/P} (x,\boldsymbol{b}_T^2)$ (right) for	the proton in the light-cone quark model as functions of $x$ at different values of $b_T$. } \label{L}
\end{figure*}

\begin{figure}[htbp]
    	\centering
    	  \includegraphics[width=\linewidth]{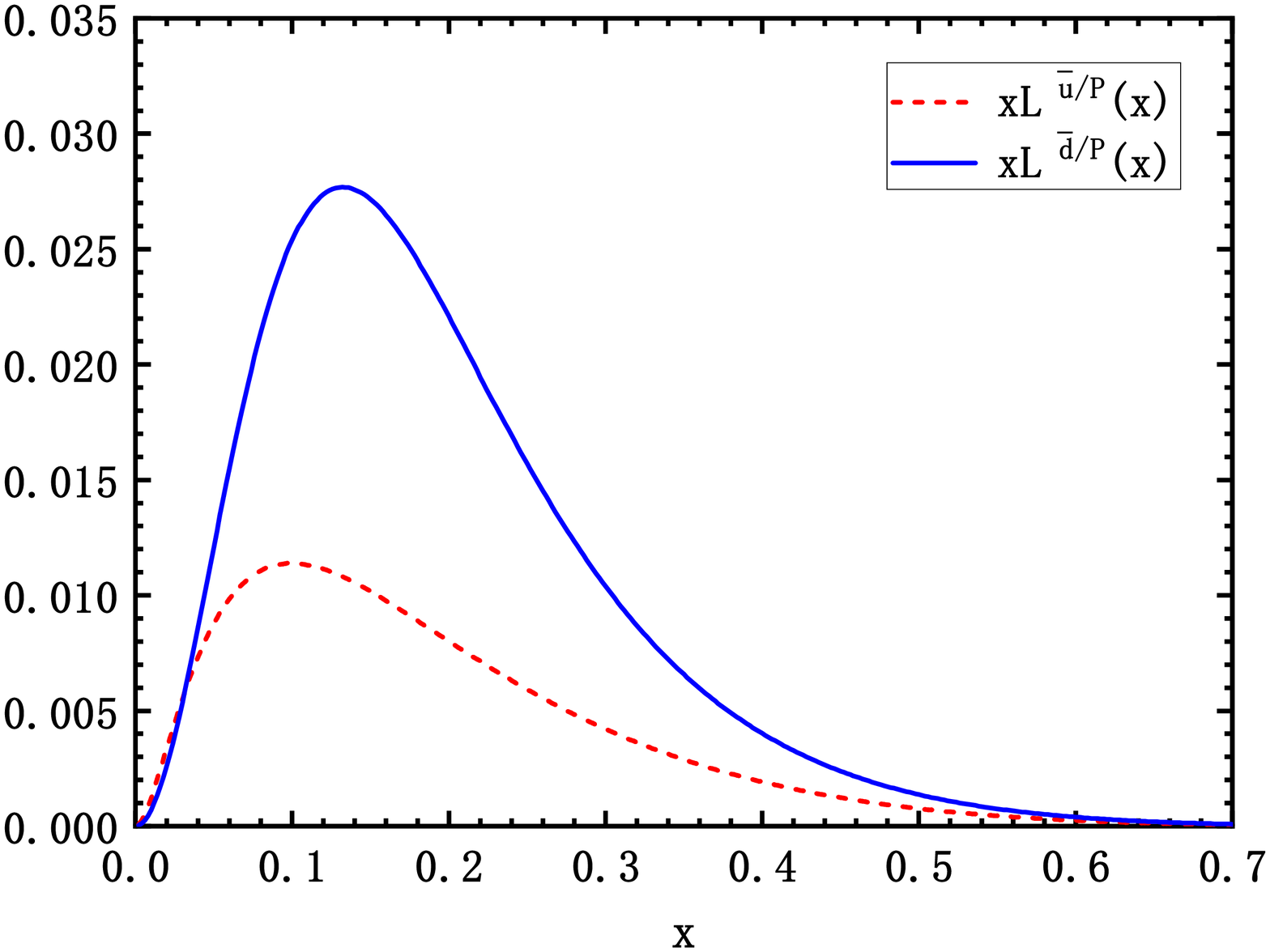}
    	\caption{ The OAM ``distribution" $L^{\bar{q}}(x)$ of $\bar{u}$ and $\bar{d}$ quarks inside the proton in the light-cone model.} \label{L-udbar}
\end{figure}

In this section, we present the numerical results for the GPDs and OAM of the sea quark $\bar{u}$ and $\bar{d}$.
For the parameters $g_1$, $g_2$, $\Lambda_{\bar{q}}$, $\Lambda_\pi$ in our model, we adopt the values shown in Table.~\ref{tab1} from Ref.~\cite{Luan:2022fjc}.

In the left and right panels of Fig.~\ref{H} and ~\ref{E}, we plot the $x$-dependence of the electric GPD $H^{\bar{q}/P}(x,0,-\boldsymbol{\Delta}_T^2)$ and the magnetic GPD $E^{\bar{q}/P}(x,0,-\boldsymbol{\Delta}_T^2)$ of the $\bar{u}$ and $\bar{d}$ quarks at different values of $\Delta_T$, respectively.
We find that in both the cases of $\bar{u}$ and $\bar{d}$, the signs of $H^{\bar{u}/P}$ and $H^{\bar{d}/P}$ are positive in the entire $x$ region, the size of $H^{\bar{d}/P}$ is larger than that of $H^{\bar{u}/P}$.
Moreover, the $x$ -dependence of $xH^{\bar{q}/P}$ varies with the change of $\Delta_T$.
To be specific, as $\Delta_T$ increases, the peak of the curves shifts from lower $\Delta_T$ to higher $\Delta_T$.

Special attention is paid to the limit of zero momentum transfer ${\Delta}_T^2=0$, since in this limit the GPD $H^{\bar{q}/P}$ reduces to the unpolarized distributions $f_{1}^{\bar{q}/P}$.
As a check, we apply the result of $xH^{\bar{q}/P}(x,0,0)$ to compare it with the unpolarized distributions $xf_{1}^{\bar{q}/P}(x)$ in Ref.~\cite{Luan:2022fjc}, we find that their numerical results are consistent.

\begin{figure*}[htbp]
    	\centering
    	\subfigure{\begin{minipage}[b]{0.45\linewidth}
    			\centering
    			\includegraphics[width=\linewidth]{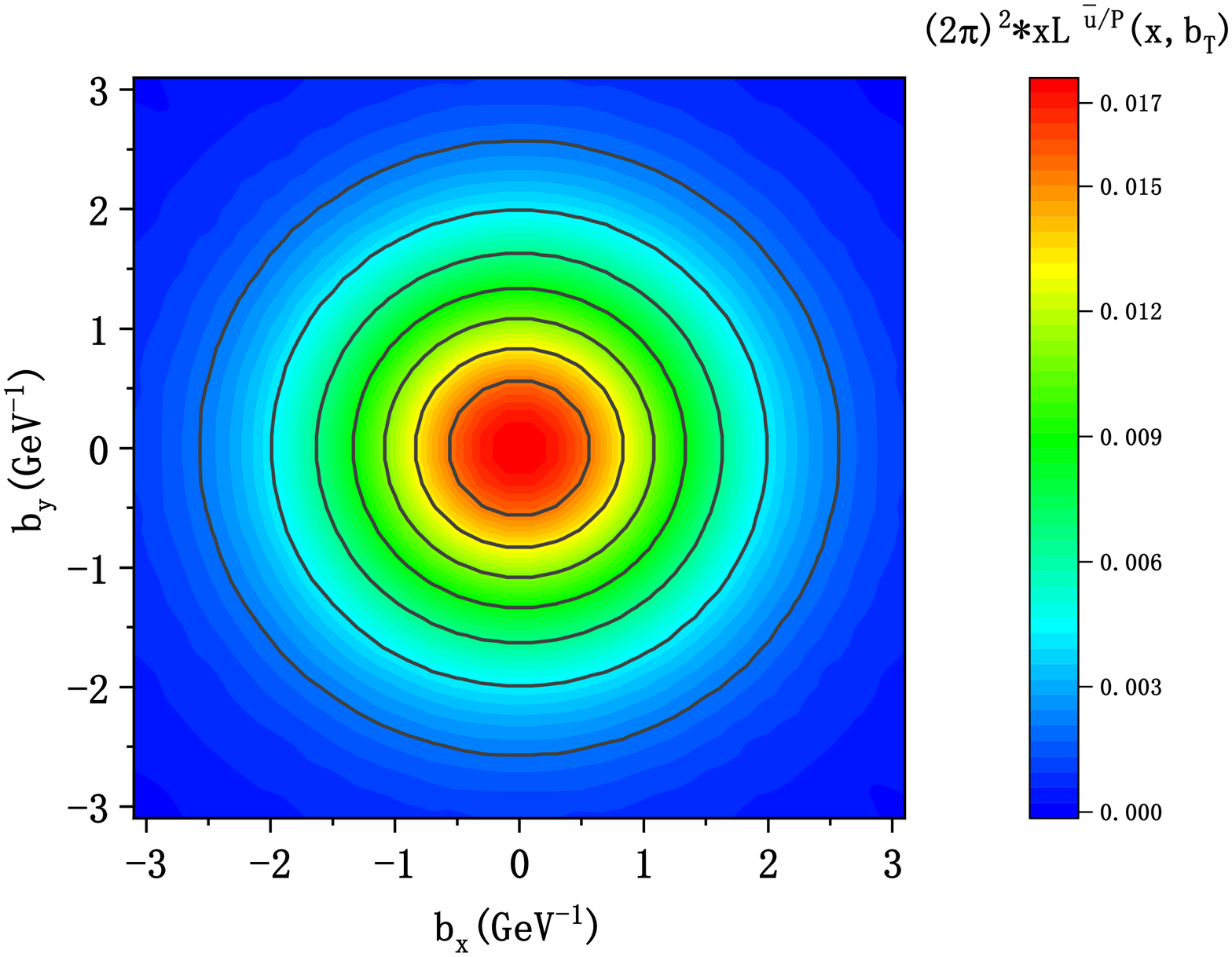}
    	\end{minipage}}
    	\subfigure{\begin{minipage}[b]{0.45\linewidth}
    			\centering
    			\includegraphics[width=\linewidth]{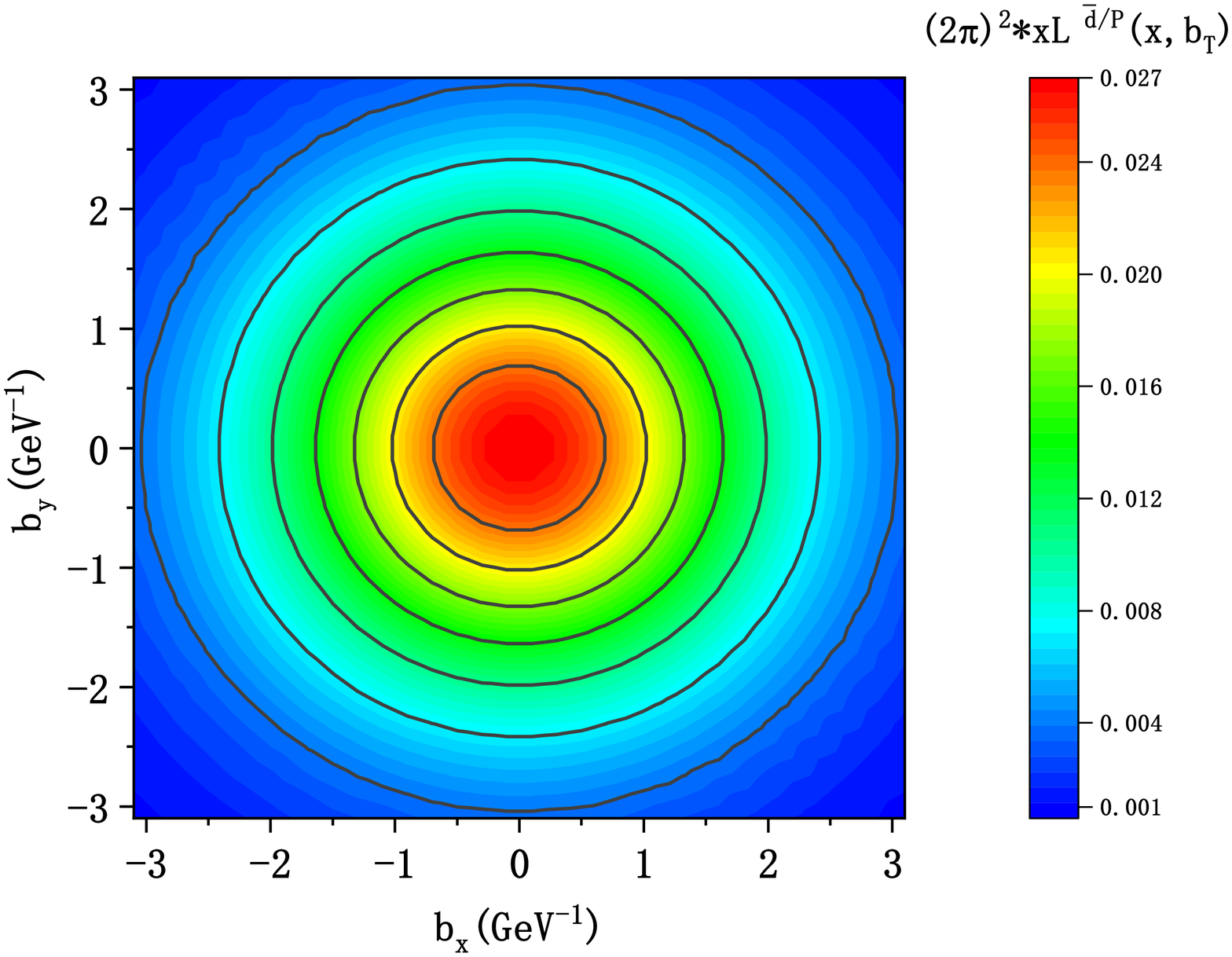}    	
       \end{minipage}}
       \caption{The profiles of the impact parameter distribution (scaled by a factor of $(2\pi)^2$) $L^{\bar{u}/P} (x,\boldsymbol{b}_T^2)$ (left) and $L^{\bar{d}/P} (x,\boldsymbol{b}_T^2)$ (right) for the proton in the light-cone model at $x = 0.3$.} \label{L-bx-by}
\end{figure*}

In Fig.~\ref{E}, the shape of $xE^{\bar{q}/P}$ is similar to that of the $xH^{\bar{q}/P}$ distribution, and its size decreases with increasing values of $\Delta_T$.
The signs of the electric GPD $xE^{\bar{u}/P}$ and $xE^{\bar{d}/P}$ are positive in the entire $x$ region, while the size of $xE^{\bar{d}/P}$ is larger than that of $xE^{\bar{u}/P}$.
    	
With the GPDs of sea quarks at hand, we can study the OAM of the $\bar{u}$ and $\bar{d}$ quark inside the proton.
The parton OAM is of great significance in understanding the nucleon spin structure.
Over the past two decades, theoretical descriptions of quark OAM have been established.
As shown in Ref.~\cite{Ji:1996ek}, the quark angular momentum can be separated into the usual quark helicity and a gauge-invariant orbital contributions $L^q$.
The latter one is related to the GPDs by Ji's sum rule:
\begin{align}
L^{\bar{q}} &=\int dx\frac{1}{2}\left\{x[H^{\bar{q}/P}(x,0,0)
+E^{\bar{q}/P}(x,0,0)]\right.\nonumber \\
&\left.-\widetilde{H}^{\bar{q}/P}(x,0,0)\right\},\label{eq:Lq}
\end{align}
Using the GPDs in our model, we calculate $L^{\bar{q}}$ and find
\begin{align}
L^{\bar{u}}=0.025,~~~~ L^{\bar{d}}=0.046.
\end{align}
This shows that both the $\bar{u}$ and $\bar{u}$ quark OAMs are positive in our model.
To understand the $x$-dependence of the OAM of sea quarks, we also keep $x$ unintegrated in Eq.~(\ref{eq:Lq}) and plot
\begin{align}
L^{\bar{q}} (x)&=\frac{1}{2}\left\{x[H^{\bar{q}/P}(x,0,0)
+E^{\bar{q}/P}(x,0,0)]\right.\nonumber \\
&\left.-\widetilde{H}^{\bar{q}/P}(x,0,0)\right\},\label{eq:Lq2}
\end{align}
in Fig.~\ref{L-udbar} for $\bar{u}$ and $\bar{d}$ quarks by the solid line and the dashed line, respectively.
We find that the OAM ``distributions" $L^{\bar{u}} (x)$ and $L^{\bar{d}} (x)$ in our model are positive in the entire $x$ region.
The size of $L^{\bar{q}/P}(x)$ is similar to the case of the GPDs, that is, the distribution of $\bar{d}$ is larger than that of $\bar{u}$.
The phenomena of these can be explained in the meson-baryon fluctuation model, that is, the possibility of the fluctuation $p\to\pi^+ n$  is larger than the possibility of the fluctuation $p\to\pi^- \Delta^{++}$.

It is also interesting to study the GPDs in the impact parameter space (transverse position space)~\cite{Burkardt:2002hr,Bondarenko:2002pp} through the Fourier transformation, as they provide a three-dimensional tomography of the nucleon.
Similarly, the impact parameter dependence of the sea quark OAM can be calculated by~\cite{Lu:2010dt}
\begin{align}
L^{\bar{q}/P} (x,\boldsymbol{b}_T^2)&=\int\frac{d^2\boldsymbol{\Delta}_T}{(2\pi)^2}
e^{-i\boldsymbol{b}_T\cdot\boldsymbol{\Delta}_T}L^{\bar{q}/P} (x,-\boldsymbol{\Delta}_T^2).
\end{align}
where is $L^{\bar{q}/P} (x,-\boldsymbol{\Delta}_T^2)$ is given by
\begin{align}
L^{\bar{q}/P}(x,-\boldsymbol{\Delta}_T^2)
&=\frac{1}{2}\{x[H^{\bar{q}/P}(x,0,-\boldsymbol{\Delta}_T^2)
+E^{\bar{q}/P}(x,0,-\boldsymbol{\Delta}_T^2)]\nonumber\\
&-\widetilde{H}^{\bar{q}/P}(x,0,-\boldsymbol{\Delta}_T^2)\},
\end{align}

In Fig.~\ref{L}, we plot the $x$-dependence of OAM distributions in the impact parameter space (scaled with a factor of $(2\pi)^2$) $L^{\bar{u}/P} (x,\boldsymbol{b}_T^2)$ (left) and $L^{\bar{d}/P} (x,\boldsymbol{b}_T^2)$ (right) at $b_T=1$ GeV$^{-1}$, 2 GeV$^{-1}$, 3 GeV$^{-1}$, respectively.
Again, the signs of $L^{\bar{q}/P} (x,\boldsymbol{b}_T^2)$ are positive in the entire $x$ region.
The size of $L^{\bar{d}/P} (x,\boldsymbol{b}_T^2)$ is lager than $L^{\bar{u}/P} (x,\boldsymbol{b}_T^2)$.
As $b_T$ increases, the peak of the curves shifts from higher $x$ to lower $x$.
In Fig.~\ref{L-bx-by}, we show the profiles of the impact parameter distributions $L^{\bar{q}/P} (x,\boldsymbol{b}_T^2)$ for the proton in the light-cone model as functions of $b_T$ at $x = 0.3$.
It is shown that the impact parameter dependence of sea quark OAM is axially symmetric.

Finally, we would like to compare our model result with the model calculation from other works.
In Ref. \cite{He:2022leb}, the sea quark GPDs are calculated using the nonlocal chiral effective theory.
In that model, the electric GPD $H$ of $\bar{u}$ and $\bar{d}$ are positive which are consistent with our result,
while $E^{\bar{u}}$ and $E^{\bar{d}}$ have opposite sign ($E^{\bar{d}}$ is positive and $E^{\bar{u}}$ is negative).
This is different from our result where the sign of $E^{\bar{u}}$ and $E^{\bar{d}}$ both are positive. In addition, the trend of the curves with the increase $\Delta_T$ at any fixed $x$ value, $H^{\bar{q}}$ and $E^{\bar{q}}$ in this model are consistent with ours.
Hopefully, the reasons for these differences can be checked through theoretical and experimental analysis in the future.

\section{CONCLUSION}\label{Sec5}

In this work, we studied the sea quark GPDs using a light-cone quark model.
We treat the Fock state of proton as a composite system formed by a pion meoson and a baryon, where the pion meson is composed of $q\bar{q}$ pair.
Using the overlap representation of LCWFs, we calculated the chiral-even GPDs of sea quark $H^{\bar{q}/P}(x,0,-\boldsymbol{\Delta}_T^2)$, $E^{\bar{q}/P}(x,0,-\boldsymbol{\Delta}_T^2)$ and $\widetilde{H}^{\bar{q}/P}(x,0,-\boldsymbol{\Delta}_T^2)$ with a convoluted form, in which
the sea quark GPDs are expressed as the convolution of the GPDs of the pion inside the proton and the antiquark GPDs inside the pion.
In the calculation, we adopted the dipole form factor to represent the coupling interacting vertices.
Numerical results show that $H^{\bar{q}/P}(x,0,-\boldsymbol{\Delta}_T^2)$, $E^{\bar{q}/P}(x,0,-\boldsymbol{\Delta}_T^2)$ are positive, while $\widetilde{H}^{\bar{q}/P}(x,0,-\boldsymbol{\Delta}_T^2)$ vanishes in the model.
The size of the $\bar{d}$ GPDs is lager than that of the $\bar{u}$ GPDs as $\Delta_T$ increases, the peak of the curves for  $H^{\bar{q}/P}(x,0,-\boldsymbol{\Delta}_T^2)$ and $E^{\bar{q}/P}(x,0,-\boldsymbol{\Delta}_T^2)$ shift from lower $\Delta_T$ to higer $\Delta_T$. The GPDs are applied to calculate the OAM distributions, showing that $L^{\bar{u}/P}(x)$ and $L^{\bar{u}/P}(x)$ are positive.
We also studied the impact parameter dependence of the sea quark OAM distribution $L^{\bar{q}/P}(x, b_T )$ which describe the position space distribution of the quark OAM at given $x$.
We find that the signs of $L^{\bar{u}/P}(x,b_T)$ and $L^{\bar{d}/P}(x,b_T)$ are positive, the value of the curve fall off monotonically with increasing $b_T$.
From the profiles of the impact parameter distribution $L^{\bar{q}/P}(x, b_T )$, we found that the impact parameter dependence of the sea quark OAM distribution is axially symmetric in the light-cone model.
We also compared our model results with the recent extraction of $H^{\bar{q}/P}(x,0,-\boldsymbol{\Delta}_T^2)$ and $E^{\bar{q}/P}(x,0,-\boldsymbol{\Delta}_T^2)$ and find some similarities and differences between them which need to be checked by future theoretical and experimental analysis.
In conclusion, our study provide useful information of the sea quark GPDs in a proton from an intuitive model.
Further study is needed in order to provide more stringent constraint on the sea quark GPDs.

\section*{Acknowledgements}
This work is partially supported by the National Natural Science Foundation of China under grant number 12150013.

\end{document}